# Biomechanical analysis of a cranial Patient Specific Implant on the interface with the bone using the Finite Element Method

J.M. Díaz[1], O. A. González-Estrada[1] and C.I. López[2]

[1] Universidad Industrial de Santander, School of Mechanical Engineering, Bucaramanga, Colombia
[2] Universidad Industrial de Santander, School of Industrial Design, Bucaramanga, Colombia

*Abstract*— **New advance technologies based on reverse engineering, design and additive manufacturing, have expanded design capabilities for biomedical applications to include Patient Specific Implants (PSI). This change in design paradigms needs advanced tools to assess the mechanical performance of the product, and simulate the impact on the patient. In this work, we perform a structural analysis on the interface of a cranial PSI under static loading conditions. Based on those simulations, we have identified the regions with high stress and strain and checked the failure criteria both in the implant and the skull. We evaluate the quality of the design of the implant and determine their response given different materials, in order to ensure optimality of the final product to be manufactured.**

## I. Introduction

Biomechanics studies the mechanical behaviour of biological systems. It helps to provide solutions to the injuries the body is not capable of healing itself, i.e., cases in which natural recovery is not possible. This could be accomplished by locating orthopaedic implants where the tissue did not heal to help with the recovery of a patient [1].

The study of these biomedical implantable devices involves areas such as materials, medicine and engineering [2]. These disciplines cannot be mutually exclusive if we want to ensure that the final product is in agreement with the expectations of the design. Moreover, it is very important to evaluate the quality of the designs before manufacturing them, in order to find any possible existing flaws, optimise the implant prototype and guarantee the patient is getting a product that is going to enhance his quality of life [3] [1].

This testing process can be done through non-destructive tests by means of numerical simulation, using finite element analysis (FEA) [4] to evaluate the implant-bone interface behaviour [5]. The model for the skull-implant assembly is considered, subjected to given boundary conditions. The structural analysis is performed in order to evaluate the stress and strain fields. Failure criteria for the materials are also considered. It is expected that critical zones are located in the interface between the skull and the implant, and that will be the main subject of this work.

## II. Methods

### A. Model identification and geometry definition

The models of the study are 3D skull reconstructions obtained with a multislice spiral CT scanner, Toshiba Aquilon, in DICOM format. Using the software Invesalius 3, with semiautomatic segmentation, they are converted to STL format. Later, they receive CAD treatment with Rhino®. Based on the reconstruction, two different implants are designed and all models imported to ANSYS 15.0.

For optimisation purposes, we removed some of the bone tissue, leaving only the bone portion of the cranial vault near to the defect, thus simplifying the model.

### B. Material models

Mechanical properties of the materials to use in the numerical models are defined. The software used in the CAD reconstruction did not allow displaying the entire Hounsfield scale [6]; therefore, the values used to represent the mechanical properties are taken from the literature. In the case of bone, Galicer [7] suggests that the value of the bone matrix density is $\rho_{mo} = 2.02 \text{ g/cm}^3$. The bone elastic constants, Young modulus and Poisson ratio, are defined by the following equations [8]:

$$E = 1763\rho 3{,}2 \text{ if } \rho > 1{,}2 \text{ g/cm}^3 \quad (1)$$

$$\upsilon = 0{,}32 \text{ if } \rho > 1{,}2 \text{ g/cm}^3 \quad (2)$$

By replacing the bone matrix density in (1), we have $E = 16725$ MPa.

Table 1 shows the Polyether ether ketone (PEEK) and titanium alloy Ti6Al4V properties, suggested by Safi [9]:

Table 1.  Young modulus $E$ and Poisson ratio $\upsilon$ for PEEK and Ti6Al4V.

| Material | $E$ [MPa] | $\upsilon$ |
|---|---|---|
| PEEK | 4000 | 0,4 |
| Ti6Al4V | 110300 | 0,36 |

*C. Contact type definition*

Many numerical models are available to define the contact conditions. For this problem we have chosen a linear contact condition, defined using Multi-Point Constraint (MPC). Linear contact models are faster to solve and less computational demanding. Extension of this work to include nonlinear models is expected. The MPC condition defines a state of no separation between the implant and the bone tissue. MPC are recommended for non-conforming contact surfaces, as it is the case for the implant-bone assembly due to features of the design introduced in the CAD model.

*D. Meshing*

Meshing is a fundamental part in the FEA, since a coarse mesh can generate results with low accuracy, whereas a fine mesh can generate a high computational cost and poor numerical conditioning. Safi [9] recommends the use of quadratic tetrahedral elements (SOLID 187).

Two nonconforming meshes were used: one for the implant and one for the skull. Both were generated with the *Patch Independent* method to overcome issues related to the STL features imported from the CAD. This process is ideal for models with complex geometries, with very sharp angles, as the algorithm ensures that the mesh is refined where needed, but maintains larger elements in areas where possible. Table 2 shows the parameters used to generate the meshes for implant 1, with a total of 529113 elements. The resulting meshes are shown in Figure 1.

Table 2. Parameters used for the mesh of implant 1.

| Parameter | Implant | Skull |
|---|---|---|
| Element's maximum size | 2 mm | 5 mm |
| Element's minimum size | 0,75 mm | 1,5 mm |
| Number of cells across a gap | 8 | 3 |
| Curvature angle | 7,5° | |
| Smooth transition | Yes | Yes |
| Number of elements | 180303 | 167450 |

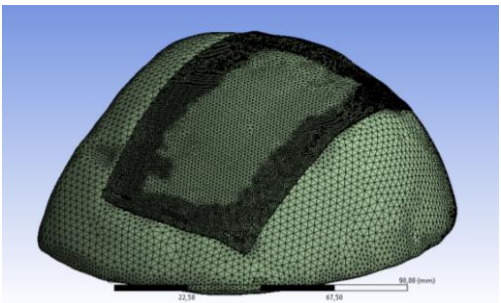

Figure 1: Final mesh for the model of the skull with implant 1.

Table 3 shows the values of the parameters used to generate the meshes for the implant 2, with a total of 681643 elements. Figure 2 shows the mesh for the implant and bone tissue.

Table 3. Parameters used for the mesh of implant 2.

| Parameters | Implant | Skull |
|---|---|---|
| Element's maximum size | 2 mm | 3 mm |
| Element's minimum size | 0,75 mm | 1 mm |
| Number of cells across a gap | 3 | 3 |
| Smooth transition | Yes | No |
| Number of elements | 199685 | 250001 |

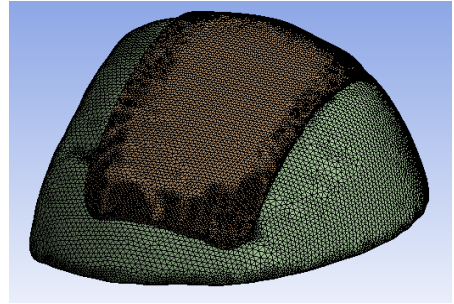

Figure 2: Final mesh for the skull with implant 2.

*E. Boundary conditions and applied loads*

In the skull part, displacement in all directions is restricted along the boundary defined by the cutting planes. For the implants, displacement is restricted in the $x$ and $y$ directions of the surfaces representing the perforations aimed for the location of the screws, thus simulating the screw clamping.

The applied load is a pressure of 1 MPa which will be applied on an circular area with a radius of 15 mm in the central zone of the implant, this is equivalent to statically applying 70 kg, as used previously in [9]. Additionally, to simulate the pressure exerted by the screws, pressure is applied around the holes aimed to hold the screws. The value is defined in terms of the torque recommended by ASTM F- 543 [6] for HA 2.0 screws:

$$P = 4T/\pi d^3 \qquad (3)$$

Where $P$ is the applied pressure, $T$ is the torque, $d$ is diameter of the head of the screw, giving a value of $P =$ 6,96 MPa.

III. Results

The quality of the implant designs will be assessed taking into account the stresses obtained for the implant and the skull, which should be below the material limits.

*A. Skull with implant 1 using PEEK*

The obtained results are analysed under the von Mises criterion, with an applied load $P = 1$ MPa. In Figure 3, we observe the obtained stress distribution. The greatest stress is located in the implant, on the region in contact with the fracture, and has a value of 96.291 MPa. This may be because the PSI implant contour adjusts tightly to the shape of the fracture, and the interference is large. It is recommended to change the CAD implant so that when the two parts come into contact interference is smaller.

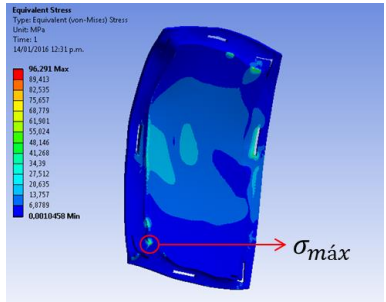

Figure 3: Stress distribution in the implant 1 using PEEK.

Figure 4 shows the strain and displacements, indicating that there are not large deformations. Structural stability is an important condition to ensure the health of the patient.

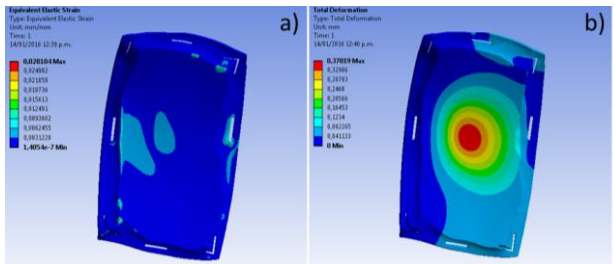

Figure 4: Strain (a) and displacements (b) in implant 1 using PEEK.

*B. Skull with implant 1 using titanium alloy (Ti6Al4V).*

The simulation is performed changing the material assigned to the implant part to titanium alloy Ti6Al4V. The maximum stress for this model is 125.83 MPa. Compared to the stress for the PEEK model, this one is slightly larger due to the higher elastic modulus. Using Ti6Al4V, it can withstand a pressure $P = 5$ MPa, with a maximum stress of 629.16 MPa, below the elastic limit of $S_y = 675$ MPa. The maximum displacement for the Ti6Al4V.is 0.01543 mm, which compared with the PEEK model is much lower. This behaviour is ideal as it is intended that the implant is deformed as little as possible once it has been placed on the patient.

However, notice that the more rigid Ti6Al4V has a stronger stress shielding effect, bad for bone resorption. Table 4 contains the results of the analysis for both implants.

Table 4: Results obtained for the model of the skull with implants 1 and 2.

| Material | Max. stress [MPa] | Max. strain | Max. displacement [mm] |
|---|---|---|---|
| *Implant 1* | | | |
| PEEK | 96,291 | 0,0281 | 0,3701 |
| Ti6Al4V Alloy | 125,83 | 0,0013 | 0,0154 |
| *Implant 2* | | | |
| PEEK | 32,575 | 0,0092 | 0,1448 |
| Ti6Al4V Alloy | 47,034 | 0,0004 | 0,0050 |

*C. Skull with implant 2 using PEEK.*

Considering the previous results we generate an enhanced implant model. We use the same boundary conditions for the implant 2. Figure 5 shows the stress distribution obtained for this model using PEEK.

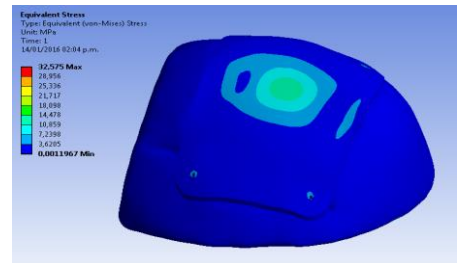

Figure 5: Stress distribution for the skull with implant 2 using PEEK.

The maximum stress is located on the face that comes into contact with the bone, with 32.575 MPa. This implant model has a smoother contour in the area, which contact the fracture. Comparing the results obtained for the implant 1 using PEEK, and given that the behaviour of the stress of this model is linear, it can resist three times more load. Figure 6 shows the strain and displacements obtained, which are much lower compared to the implant 1 using PEEK. The maximum displacement has a value of 0.1448 mm.

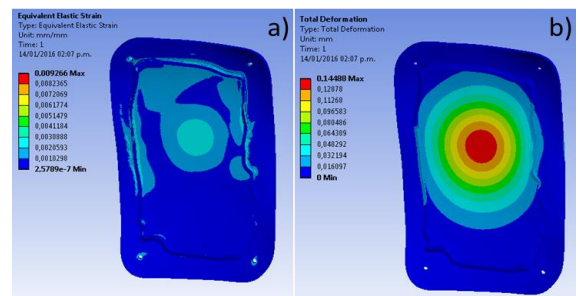

Figure 6: Strain (a) and (b) displacements in implant 2 using PEEK.

*D. Skull with implant 2 using titanium alloy (Ti6Al4V).*

Simulation is performed using the titanium alloy Ti6Al4V. Figure 7 shows the stress distribution for this case.

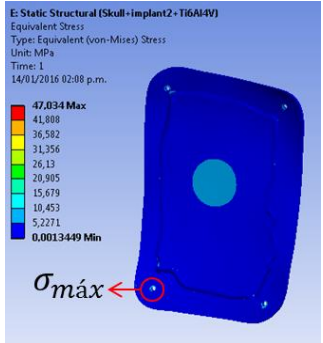

Figure 7: Stress distribution in the implant 2 using Ti6Al4V.

Maximum stress is located in the holes provided to secure the implant with screws and have a value of 47 MPa. Different behaviour from the previous model can be observed because the titanium alloy is more rigid than PEEK and the contour that gets in contact with the bone is practically not deformed so the holes act as stress concentrators. Comparing this value with the obtained in implant 1 using Ti6Al4V, implant 2 resists a load three times higher. Figure 8 shows the strain and displacements respectively, with a max. displacement of 0.005 mm.

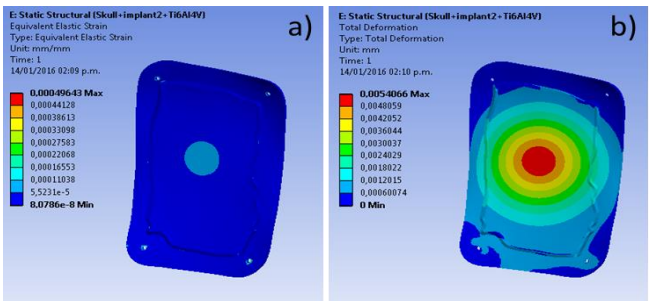

Figure 8: Strain (a) and (b) displacements in implant 2 using Ti6Al4V.

## IV. CONCLUSIONS

For this work it was assumed that the bone behaves as an isotropic linear elastic material, which is an approximation to simplify the analysis. Meshing is a key part of the analysis and requires great effort in the preprocessing stage. Manual interaction is still required to produce good quality meshes for such type of problems. A linear contact model was considered, with prestress induced by the screws.

In the analysis performed, implant 2 is superior to implant 1 model, primarily because implant 2 is a redesign of the implant 1, in which the geometry is redefined to improve the contour. PEEK exhibits lower values of stress that the more rigid Ti6Al4V, which is in agreement with the known stress shielding effect of Titanium alloys. We show the feasibility of virtual prototyping for modelling PSI implants.

Extension of this work could include nonlinear models for contact conditions, and more complex material models, including varying bone density obtained from the scans.

## ACKNOWLEDGMENT

This work was supported in part by VIE-UIS under Code No. 1361.

Author:   O. A. González-Estrada; C.I. López2
Institute: Universidad Industrial de Santander
Street:   Cra 27 street 9
City:     Bucaramanga
Country:  Colombia
Email:    agonzale@uis.edu.co; clalogu@uis.edu.co